\newif\ifcheckpagelimits
\checkpagelimitstrue
\checkpagelimitsfalse

\ifcheckpagelimits

 \documentclass[nofootinbib,prl,aps,twocolumn,showpacs,showkeys,amsmath,amssymb,superscriptaddress,final,reprint,floatfix,longbibliography]{revtex4-1}

 \newcommand{\todo}[1]{}
\else
 \documentclass[prl,aps,twocolumn,showpacs,showkeys,%
 amsmath,amssymb,superscriptaddress,final,reprint,floatfix,longbibliography]{revtex4-1}
 \newcommand{\todo}[1]{{\pdfmargincomment[icon=Note,color=pink]{#1}}}
\fi

\usepackage{lineno}
\usepackage{subfigure}
\usepackage{dcolumn}
\usepackage{amsmath,amssymb}

\usepackage{bm}
\usepackage{color}
\usepackage{overpic}
\usepackage{latexsym}
\usepackage{epstopdf}
\usepackage[english]{babel}
\usepackage{latexsym}
\usepackage{stmaryrd}

\definecolor{mygrey}{gray}{0.35}
\definecolor{myblue}{rgb}{0.2,0.2,0.8}
\definecolor{myzard}{cmyk}{0,0,0.05,0}
\definecolor{mywhite}{rgb}{1,1,1}
\definecolor{myred}{rgb}{1,0.,0.3}

\usepackage[colorlinks=true,citecolor=myblue,linkcolor=myred,urlcolor=myblue]{hyperref}

 \def\ee{\mathord{\rm e}}

\renewcommand{\ee}{{\rm e}}
\def\beq{\begin{equation}}
\def\eeq{\end{equation}}

\usepackage{graphicx}
\usepackage{soul}
\usepackage{epstopdf}

\usepackage{dsfont}
\newcommand{\simleq}{\; \raisebox{-0.4ex}{\tiny$\stackrel
{{\textstyle<}}{\sim}$}\;}
\newcommand{\ket}[1]{\left\vert #1 \right\rangle}

\def \ket#1{|#1\rangle}

\def \be{\begin{equation}}
\def \ee{\end{equation}}
\def \ba{\begin{array}}
\def \ea{\end{array}}
\def \bea{\begin{eqnarray}}
\def \eea{\end{eqnarray}}

\def\O {\Omega}

\renewcommand{\phi}{\varphi}

\usepackage{tabularx,ragged2e,booktabs}
\newcolumntype{C}[1]{>{\Centering}m{#1}}

\begin{document}

\title{Scheme for detection of single-molecule radical pair reaction using spin in diamond}

\author{Haibin Liu}
\affiliation{School of Physics $\&$ Center for Quantum Optical Science, Huazhong University of Science and Technology, Wuhan 430074, P. R. China}

\author{Martin B. Plenio}
\affiliation{Institut f\"{u}r Theoretische Physik \& IQST, Albert-Einstein Allee 11, Universit\"{a}t Ulm, 89069 Ulm, Germany}

\author{Jianming Cai}
\email{jianmingcai@hust.edu.cn}
\affiliation{School of Physics $\&$ Center for Quantum Optical Science, Huazhong University of Science and Technology, Wuhan 430074, P. R. China}

\begin{abstract}
The radical pair reaction underlies the magnetic field sensitivity of chemical reactions 
and is suggested to play an important role in both chemistry and biology. Current experimental 
evidence is based on ensemble measurements, however, the ability to probe the radical pair reaction at the
single molecule level would provide valuable information concerning its role in important
biological processes. Here, we propose a scheme to detect the charge recombination rate in a
radical pair reaction under ambient conditions by using single nitrogen-vacancy center spin in diamond.
We demonstrate theoretically that it is possible to detect the effect of  the geomagnetic
field on the radical pair reaction and propose the present scheme as a possible
hybrid model chemical compass.
\end{abstract}


\date{\today}

\maketitle

{\it Introduction.---} Radicals are frequently involved in chemical reactions, and play a
crucial role in biological processes such as cell function, magnetic sensing and diseases \cite{Droe_02_PR,Sch_78_ZPC,Rit_00_BJ, Joh_05_NRN,Rod_09_PNASU,Hal_15_xx}. The available methods to measure the properties of  reactions involving radicals include electron spin resonance \cite{Fuk_96_M}, nuclear magnetic resonance \cite{Haw_96_M} and high performance liquid chromatography \cite{Zha_05_PNASU}. These methods require large ensembles to derive a detectable signal and can therefore only provide 
average information about the reaction. Detection techniques at single-molecule level 
would provide unparalleled access to individual reactions and their pathways providing information 
that is hidden in the ensemble average. An important potential target for single molecule studies 
is the magnetic field dependent radical pair reaction that provides one plausible mechanism for 
biological magnetoreception \cite{Sch_78_ZPC,Rit_00_BJ,Rod_09_PNASU}. The direct observation of 
the reaction kinetics at single-molecule level would provide 
unambiguous insights into the involved quantum coherent spin dynamics and its non-trivial role in bio-magnetic sensing \cite{Cai_10_PRL,Gau_11_PRL,Cai_12_PRA,Hog_12_PRL,Cai_13_PRL}.

Recently, high-sensitivity nanoscale sensors based on nitrogen-vacancy (NV) center
spin in diamond have been actively pursued \cite{Sch14,WuRev16}. The nitrogen-vacancy center spin has
superior quantum properties, including long spin coherence time under ambient conditions, non-bleach
photoluminescence, and high sensitivity to external signals. These properties make the
single nitrogen-vacancy center spin sensor an appealing and promising candidate for quantum sensing
of various physical parameters, such as magnetic
field \cite{Maz_08_NL,Bal_08_NL,Tay_08_NP}, electric field \cite{Dol_11_NP,Dol_14_PRL}, single
electron and nuclear spin \cite{Zhao11,Boss16,Copper14,Gri_13_NP,Shi_15_S,Muel_14_NC,DeV_15_NN},
and temperature \cite{Kuc_13_NL,Toy_13_PNASU,Neu_13_NL}. Moreover, the biocompatibility and
chemical inertness make it feasible to perform sensitive  measurement in biological systems. These properties make the NV center spin sensor a promising candidate to detect the radical pair reaction with the potential to open up a new way to study its implications in chemistry and biology.

In this work, we propose a method for using a single NV center spin sensor to detect charge
separation and measure the charge recombination rate of the radical pair at the single molecule
level under ambient conditions. The present approach works for the scenario of radical pair reaction
with fixed radical-radical distance and non-zero electric dipole, e.g. formed by electron transfer
from one uncharged molecule to another uncharged molecule and with no change in the protonation
state. For simplicity, we assume that the depth of the NV centre is above 5nm so that it is sufficiently far away from the radical pair to render the influence of the spin-spin (exchange or dipolar) coupling between them on the spin dynamics in both the radical pair and the NV centre negligible. The formation
of  the radical pair induces an electric dipole which affects the energy levels of the NV
center electron spin and can be used to monitor the event of charge separation and recombination. This
provides the possibility to extract information on the reaction kinetics, such as the charge recombination
rate, by measuring the dynamic evolution of an NV center spin sensor. By careful analysis, we estimate
that the measurement sensitivity of the charge recombination rate can achieve $\eta_{\text{op}}=0.54$ kHz Hz$^{-\frac{1}{2}}$ using a single NV center at a depth of 5nm. We demonstrate that this sensitivity
is sufficient to detect the response of a radical pair to the orientation of  the geomagnetic field,
owing to the anisotropic behavior of the effective charge recombination rate. Such a hybrid system may
help to construct a model chemical compass \cite{Hore08} working for geomagnetic field. The extension
to nanodiamond spin sensor may provide a possible tool to detect the radicals e.g.  in living cells \cite{Liam11,Kuc_13_NL}.

{\it Single-molecule radical pair reaction sensing.---} Radical pairs are involved in many chemical
and biological processes. Each radical has an unpaired electron that is coupled with its surrounding
nuclei and an external magnetic field $\vec{B}$ (if applied) via the Hamiltonian \cite{Ste_89_CR}
\begin{equation}
    H_{RP}=g_{e}\mu_{B}\sum_{k}\vec{B}\cdot\vec{S}_{k}+\sum_{k,i}\vec{S_{k}}\cdot A_{i}^{k}\cdot\vec{I}_{i}^{k},
\end{equation}
where $\vec{S}_{k}(k=1, 2)$ are the spin operators for two radicals, $\vec{I}_{i}^{k}$ is the spin
operator for nucleus $i$, and $A_{i}^{k}$ 
is the corresponding hyperfine interaction tensor, $g_{e}$ is the electron g factor, $\mu_{B}$ is 
the Bohr magneton. For simplicity, the dipole-dipole and exchange interaction between two radicals 
are neglected, which is valid when the radical-radical distance is sufficiently large \cite{Efi_08_BJ}.

\begin{figure}[t]
\begin{centering}
\includegraphics[width=0.9\columnwidth]{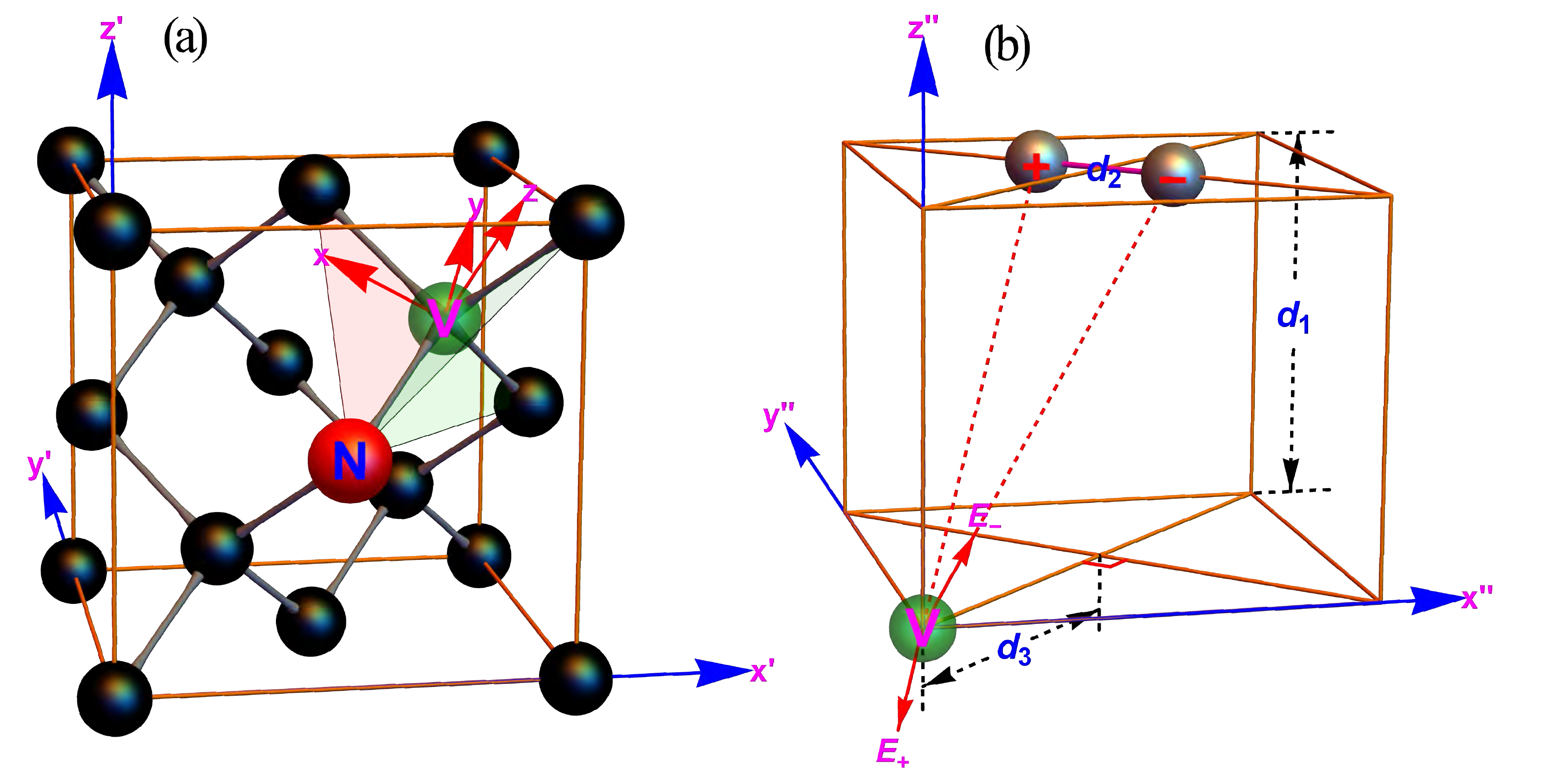}
\par\end{centering}
\caption{(Color online) {Single-molecule radical pair reaction sensing using single spin in diamond.}
\textbf{(a)} shows a unit cell of diamond containing a nitrogen-vacancy center. The nitrogen (N)--vacancy
(V) pair orientates along the $\hat{z}$ axis that is the trigonal symmetry axis of the center, and three
mirror planes are shown in green and red (the plane that contains the $\hat{x}$ axis is in red). \textbf{(b)}
The interaction between the NV center spin and the radical pair arises from the electric dipole of the
radical pair in the charge separated state. \label{fig:NV-cell} }
\end{figure}

Usually, the two unpaired radical electrons are formed upon excitation by light in a spin entangled
state, i.e., they begin either in a singlet $\left|s\right\rangle $ or triplet state $\left|t_{i}
\right\rangle $ ($i=0,+,-$). To avoid the influence of light required to prepare and read out the NV center's spin state on the radical pair, typically achieved by a 532nm laser, it is preferable to choose radical pair that is sensitive to light of a different wavelength. It is important to note that the subsequent reaction channels are
spin state dependent. Therefore, an external magnetic field can affect the chemical reaction product,
which hus forms the basis of a magnetic chemical compass \cite{Sch_78_ZPC,Rit_00_BJ,Joh_05_NRN,Rod_09_PNASU}.
The spin-selective recombination into the singlet and triplet product states can be described by the superoperators
$\mathcal{D}_{u}(\rho)=2L_{u}\rho L_{u}^{\dagger}-L_{u}^{\dagger}L_{u}\rho-\rho L_{u}^{\dagger}L_{u}$, with
$u=s,t_{0},t_{\pm}$ representing the singlet and triplet recombination channels. The Lindblad
operators are $L_{s}=\left|s,G\right\rangle \left\langle s,E\right|$, $L_{0}=\left|t_{0},G\right\rangle
\left\langle t_{0},E\right|$, $L_{-}=\left|t_{-},G\right\rangle \left\langle t_{-},E\right|$, $L_{+}=\left|t_{+},G\right\rangle \left\langle t_{+},E\right|$, $\left|E\right\rangle $ and $\left|G\right\rangle $ indicate that the radical pair is in the charge separated and recombined state. Thus the radical pair's
dynamics can be described by $\dot{\rho}_{RP}=-i\left[H_{RP},\rho_{RP}\right]+\sum_{u}k_{u}\mathcal{D}_{u}(\rho_{RP})/2,$ where $k_{u}$
are the chemical reaction rates for different spin states. This master equation is equivalent to the
conventional phenomenological master equation \cite{Kom_09_PRE,Jon_10_CPL}, and allows us to
generalize to the scenario of interaction with an NV center spin.

If the radical pair is in the charge separated state $\left|E\right\rangle $, the two electrons are
unpaired, which form an electric dipole due to their spatial separation and produce a non-zero
electric field acting on an NV center spin. While the radical pair is in the charge
recombined state $\left|G\right\rangle $, the two electrons are paired, the electric field vanishes.
Our scheme is based on the response of an NV center spin to an external electric field as produced by
the radical pair. The ground state manifold of a NV center is spin-1 triplet with a zero-field energy
splitting of $D/(2\pi)\thickapprox 2.87$ GHz between its $m_{s}=\pm1$ and $m_{s}=0$ sub-levels at room temperature.
The dynamics of an NV center spin is governed by the following Hamiltonian \cite{Doh_12_PRB,Dol_14_PRL} as
\begin{eqnarray}
    H_{NV} & = & (D+k_{\parallel}E_{z})(S_{z}^{2}-2/3)+g_{e}\mu_{B}\vec{B}\cdot\vec{S}\nonumber \\
    &  & -k_{\perp}E_{x}(S_{x}^{2}-S_{y}^{2})+k_{\perp}E_{y}(S_{x}S_{y}+S_{y}S_{x}),\label{eq:Hnv}
\end{eqnarray}
where $\vec{S}$ are the NV center spin operator, $\vec{B}$ and $\vec{E}$ are the magnetic and electric
field respectively, $k_{\parallel}/(2\pi)=0.0035(2)$ Hz m/V and $k_{\perp}/(2\pi)=0.17(3)$ Hz m/V are the electric
susceptibility parameters along the direction that is in parallel and perpendicular to the NV axis \cite{Van_90_CPL}. We denote $\hat{z}$ axis as
the NV axis and $\hat{x}$ axis as the axis lying in the NV center's mirror planes, see Fig.\ref{fig:NV-cell}(a).

\begin{figure*}
\begin{centering}
\includegraphics[width=1\textwidth]{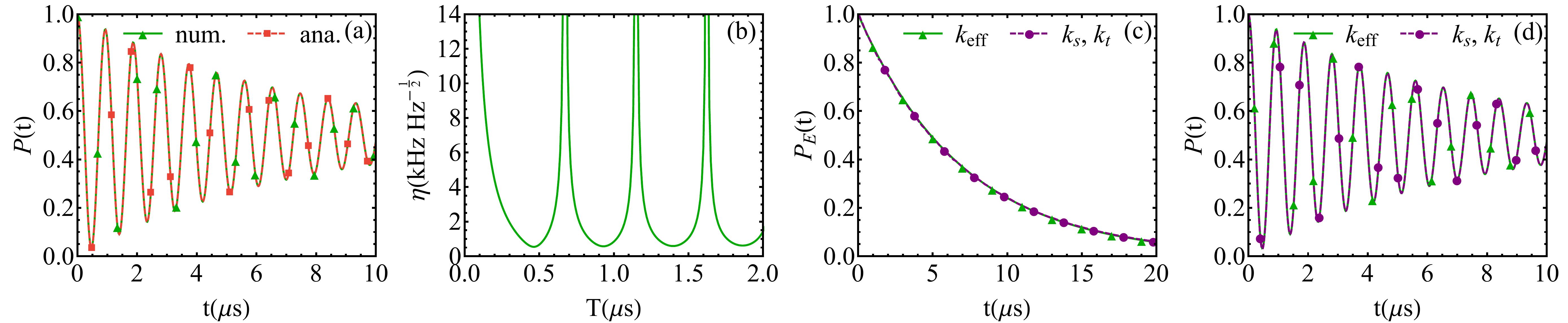}
\par\end{centering}
\caption{(Color online) {Quantum sensing of single-molecule radical pair reaction of Flavin radical ($\text{FH}{}^{\bullet-}$) and tryptophan radical ($\text{W}{}^{\bullet+}$)}. \textbf{(a)} The probability $P(t)$ that the NV center spin is in the state $\left|1\right\rangle $ as a function of the evolution time $t$. The red dashed line is analytic result and the green one is obtained from numerical simulation. \textbf{(b)} The sensitivity for the measurement of the charge recombination rate $k$ as a function of the interrogation time $T$. The best sensitivity $\eta_{\text{op}}=0.54$ kHz Hz$^{-\frac{1}{2}}$ is achieved with $T=0.46\mu s$. \textbf{(c)} and \textbf{(d)} show the probabilities that the radical pair remains in the charge separated state $\left|E\right\rangle $ and the NV center is in the state $\left|1\right\rangle $ respectively as a function of the evolution time $t$. The Green lines are the results of $k_s=k_t=k_{\text{eff}}$, the purple dashed lines are that of $k_{s}$ and $k_{t}$ with different values. The parameters are $B_{0}=0.05$ mT, $\theta=\frac{\pi}{2}$, $\phi=2.0$, $k_{s}=0.02$ MHz, $k_{t}=0.2$ MHz, and $k=k_{\text{eff}}=0.1425$ MHz. For simplicity, we consider H$^{6}$ with the dominant anisotropic hyperfine in the $\text{FH}{}^{\bullet-}$ radical \cite{Cin_03_CP}. The radical pair molecule is $d_{2}=2$ nm apart as located on the diamond crystal surface with a horizontal distance about $d_{3}=4$ nm from the NV center, and the depth of the NV center is $d_1=5$ nm, see Fig. \ref{fig:NV-cell}(a) and (b).\label{fig:Pt-eta} }
\end{figure*}

As an example, we consider a $\left\langle 001\right\rangle $ surface diamond crystal, and 
a shallow implanted NV center, e.g. at a depth of $d_{1}$ around $5$ nm from the diamond surface 
\cite{Muel_14_NC}, the axis of which is aligned along the $\left[111\right]$ crystal axis. To facilitate the coherent rotation between $\ket{{+1}}$ and $\ket{{-1}}$,  it requires $k_\perp E_\perp  \geq g \mu_B B_z $, thus we apply a weak magnetic field $\vec{B}=B_{0}(\sin\theta\cos\phi,\sin\theta\sin\phi,\cos\theta)$ that is perpendicular to the NV axis, i.e.,
$\theta=\frac{\pi}{2}$. The magnetic field would affect the spin dynamics of the radical pair reaction as we
tune its direction around the NV axis. In our protocol, we initially prepare the NV center spin in the state $\left|1\right\rangle \equiv\vert {m_{s}=+1 }\rangle$,  which may be achieved either via frequency selection of the transition $\left|0\right\rangle \rightarrow \vert 1\rangle$ or, especially in small magnetic fields, via application of a circularly polarized microwave field \cite{Alegre07,London14}. The radical pair is assumed to be created in the singlet state $\left|s\right\rangle $ as excited by a short laser pulse, and the relevant nuclear spins interacting with the radicals are in the thermal equilibrium state, which is described by a density matrix as $\bigotimes_{j}I_{j}/d_j$ ($j$ indicates the $j$-th nucleus and $d_j$ is the dimension of the corresponding Hilbert space) under ambient conditions. For simplicity, we polarize the $^{14}$N nuclear spin of an NV center to the state $\left| 0\right\rangle $ so that its hyperfine interaction with the NV center spin is effectively eliminated. We remark that the polarization of $^{14}$N nuclear spin using an NV center has been achieved in several experiments, see e.g. \cite{Jac_09_PRL,Sme_09_PRA}. The initial state of the total system of the NV center spin and the radical pair is thus written as follows
\begin{equation}
    \rho(0)=\left|1\right\rangle \left\langle 1\right|\otimes\left|s\right\rangle \left\langle s\right|\otimes\left(\bigotimes_{j}\frac{I_{j}}{d_j}\right)\otimes\left|E\right\rangle \left\langle E\right|.
\end{equation}
The dynamics of the total system can be described by the following Lindblad-type quantum master equation
\begin{eqnarray}
    \dot{\rho} & = & -i\left[H,\rho\right]+\frac{k_{s}}{2}\left\{ 2L_{s}\rho L_{s}^{\dagger}-L_{s}^{\dagger}L_{s}\rho-\rho L_{s}^{\dagger}L_{s}\right\} \nonumber \\
    &  & +\frac{k_{t}}{2}\sum_{i=t_{0},t_{\pm}}\left\{ 2L_{i}\rho L_{i}^{\dagger}-L_{i}^{\dagger}L_{i}\rho-\rho L_{i}^{\dagger}L_{i}\right\} ,\label{eq:ME}
\end{eqnarray}
with the total Hamiltonian as
\begin{equation}
H=(H_{NV}+H_{RP})\otimes\left|E\right\rangle \left\langle E\right|+H_{NV}\vert_{ E=0 }\otimes\left|G\right\rangle \left\langle G\right|,
\end{equation}
where $H_{NV}\vert_{ E=0 }$ represents the Hamiltonian of the NV center spin as in Eq.(\ref{eq:Hnv}) with a zero electric field.

The signal that we measure from the NV center spin sensor is the population of state
$\left|1\right\rangle $ after an evolution of time $T$. To demonstrate the basic idea, we 
start from the simple case of $k_{s}=k_{t}\equiv k$. The signal $P(t)$ is solved analytically 
from the master equation as in Eq.\eqref{eq:ME}. After tracing out the degrees of freedom of 
the radical pair, we obtain the following two differential equations for the NV center spin 
reduced state operator corresponding to the subspace of $P_{E}=\left|E\right\rangle 
\left\langle E\right|$ and $P_{G}=\left|G\right\rangle \left\langle G\right|$ as
\begin{equation}
\left\{ \begin{array}{l}
\dot{\rho}_{E}=-i[H_{E},\rho_{E}]-k\rho_{E}\\
\dot{\rho}_{G}=-i[H_{G},\rho_{G}]+k\rho_{E}
\end{array}\right.,\label{eq:rho-eg}
\end{equation}
where $H_{E}=H_{NV}+H_{RP}$, $H_{G}=H_{NV}\vert_{ E=0 }$, $\rho_{E}=Tr_{RP}(P_{E}\rho P_{E})$, $\rho_{G}=Tr_{RP}(P_{G}\rho P_{G})$ and their initial conditions are $\rho_{E}(0)=\left|1\right\rangle \left\langle 1\right|$ and $\rho_{G}(0)=0$. We find the solution of $\rho_{E}$ as follows
\begin{equation}
\rho_{E}(t)=e^{-iH_{E}t}\rho_{E}(0)e^{iH_{E}t}e^{-kt}.\label{eq:rho-e}
\end{equation}
In the present scenario, we consider a weak external magnetic field (for example geomagnetic field),
namely $g_{e}\mu_{B}B_{0} \ll D$, and in addition $k\ll k_{\perp}E_{\perp}$, which can be satisfied
for a shallow implanted NV center \cite{Ofori12}. The signal $P(t)$ can be separated into two parts
of contribution $P ^{E}(t)=\text{Tr}\left[\rho_{E}(t)\left|1\right\rangle \left\langle 1\right|\right]$
and $P ^{G}(t)\simeq k\int_{0}^{t}P ^{E}(\tau)d\tau$ as follows \cite{SI}
\begin{equation}
    P ^{E}(t)\simeq\frac{1+\cos(\Omega t)}{2}e^{-kt}, \quad P ^{G}(t)\simeq\frac{1-e^{-kt}}{2},
\end{equation}
where $\Omega\simeq2k_{\perp}E_{\perp}$ is the Rabi frequency due to the electric field produced by
the radical pair. Therefore, the signal of the NV center spin sensor is given by
\begin{equation}
    P (t)=P ^{E}(t)+P ^{G}(t)\simeq\frac{1+\cos(\Omega t)e^{-kt}}{2}.\label{eq:pt}
\end{equation}
We compare the above analytic result with the exact numerical simulation in Fig.\ref{fig:Pt-eta}(a)
which shows good agreement. It can be seen that the signal oscillates with a decaying envelop at a
rate that equals to the charge recombination rate of the radical pair reaction. Therefore, the signal
from the NV center spin sensor allows us to extract the relevant information. The measurement sensitivity
for the estimation of the charge recombination rate $k$ from the signal $P(t)$ is given by \cite{Wineland1992}
\begin{equation}
    \eta=\frac{\Delta P (T)}{\left|\partial P (T)/\partial k\right|}\sqrt{T},
\end{equation}
where $\Delta P (T)$ is the shot noise limited measurement uncertainty, and $T$ is the interrogation
time. The result is shown in Fig.\ref{fig:Pt-eta}(b), from which the best achievable sensitivity is
estimated to be $\eta_{\text{op}}\simeq0.54$ kHz Hz$^{-\frac{1}{2}}$ with $T\simeq  {\pi}/{\Omega} =0.46\mu s$
using realistic parameters for the NV center. Note that it is preferable to have the applied magnetic
field is perpendicular to the NV axis, as was assumed in our scheme. The effect of
the transverse component of the magnetic field can be cancelled using decoupling pulses (see Supplementary Information), which has been well developed
for NV centers. 

For the general case of $k_{s}\neq k_{t}$, according to the probability that the radical pair remains
in the charge separated state $\left|E\right\rangle $, denoted as $P_{E}(t)$, we find that it is possible
to fit an effective charge recombination rate $k_{\text{eff}}$ such that $P_{E}(t)\simeq\exp(-k_{\text{eff}}t)$,
see e.g. Fig.\ref{fig:Pt-eta}(c). It can be seen from Fig.\ref{fig:Pt-eta}(c) and (d) that the fitted
value of $k_{\text{eff}}$ lead to almost the same signal of the NV center spin sensor $P (t)$ as the
real spin-selective rate parameters $k_{s}$ and $k_{t}$. Our proposal thus offers a way to determine
the effective charge recombination rate in such scenarios. The difference between $k_{s}$ and $k_{t}$
leads to the magnetic field response of the radical pair reaction. In particular, the orientation of
the magnetic field would affect the radical pair lifetime, namely the effective charge recombination
rate $k_{\text{eff}}$. As an example, it can be seen that a magnetic field as weak as the geomagnetic
field would cause about $10\%$ change in the effective charge recombination rate of a radical pair reaction
of Flavin radical ($\text{FH}{}^{\bullet-}$) and tryptophan radical ($\text{W}{}^{\bullet+}$), see Fig.S1
in Supplementary Information, which lies well within the measurement sensitivity achievable by the single
NV center spin sensor. This makes it feasible to demonstrate the working principle of a chemical compass
based on radical pair mechanism at single-molecule level. It is important to note that in the present
scheme, the best achievable sensitivity (to the effective reaction rate) is not equivalent to the 
best achievable sensitivity to the magnetic field, and is hardly dependent on the multinuclear 
condition where radical pair interacts with more than one nucleus. On the other hand, the change in the effective charge recombination rate indeed is 
dependent on the number and species of nuclear spins in the radical pair. If such a change is 
larger than the best achievable sensitivity, then it is possible to detect the magnetic field effect 
on the radical pair.

\begin{figure}[t]
\begin{centering}
\includegraphics[width=1\columnwidth]{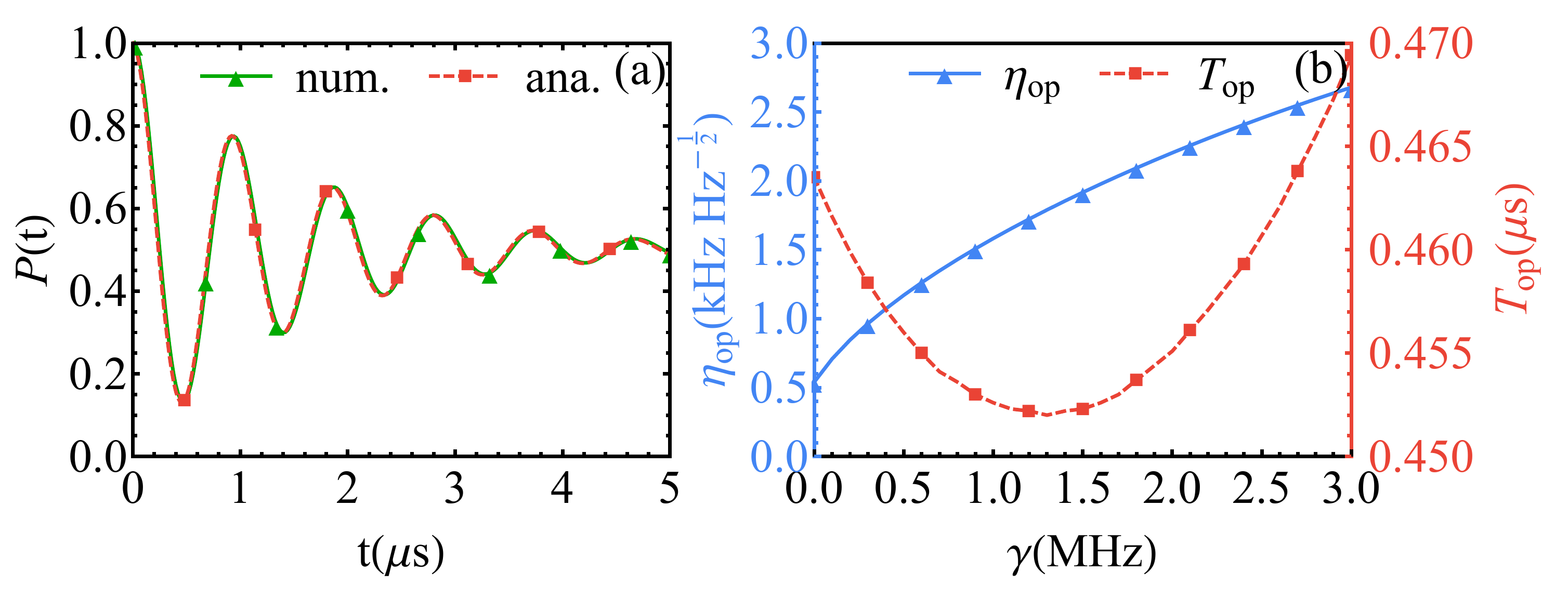}
\par\end{centering}
\caption{(Color online) {Noise effect on the measurement sensitivity.} \textbf{(a)} The probability that the NV center spin sensor is in the state $\left|1\right\rangle $ evolves with time $t$ in a dephasing environment with a dephasing rate $\gamma=0.5$ MHz. The red dashed line is the analytical result as compared with the numerical simulation (green). \textbf{(b)} The influence of the dephasing rate $\gamma$ on the best achievable sensitivity $\eta_{\text{op}}$ and the optimal interrogation time $T_{\text{op}}$. The parameters are $B_{0}=0.05$ mT, $\theta=\frac{\pi}{2}$, $\phi=2.0$ and $k=k_{\text{eff}}=0.1425$ MHz. The location of the NV center and the radical pair molecule is the same as Fig.\ref{fig:Pt-eta}.\label{fig:Pt-eta-de} }
\end{figure}

{\it Analysis of noise effect.---} For an NV center spin in diamond, the main effect of magnetic field noise is
dephasing. To demonstrate the achievable measurement sensitivity when taking into account realistic
noise levels, without loss of generality, we describe noise as pure dephasing by the Lindblad operator
$\mathcal{L}_{d}=\sqrt{\gamma}S_{z}$, where $\gamma$ is the dephasing rate, $S_{z}$ is the $\hat{z}$-component
of the NV center spin operator. The dynamics of the total system of the NV center and the radical pair
can be described by the following master equation
\begin{eqnarray}
\dot{\rho} & = & \text{right-hand side of Eq.\eqref{eq:ME}}\nonumber \\
 &  & + \gamma \left\{ S_{z}\rho S_{z}-\frac{1}{2}S_{z}S_{z}\rho-\frac{1}{2}\rho S_{z}S_{z}\right\} .\label{eq:ME-dephasing}
\end{eqnarray}
Under the additional assumption that $\gamma\ll k_{\perp}E_{\perp}$, and considering $\vec{B}$ that is perpendicular to the NV axis, we get the analytical solutions of $P ^{E}(t)\simeq\left[1+e^{-\gamma t}\cos(\Omega t)\right]e^{-kt}/2$, $P ^{G}(t)\simeq\left(1-e^{-kt}\right)/2$ \cite{SI}. Thus, the signal of the NV center spin sensor is written as
\begin{equation}
P (t)=P ^{E}(t)+P ^{G}(t)\simeq\frac{1+\cos(\Omega t)e^{-(k+\gamma)t}}{2}.\label{eq:pt-de}
\end{equation}
The above analytical result agrees well with the numerical calculation, see Fig.\ref{fig:Pt-eta-de}(a). The optimal sensitivity for the measurement of the charge recombination rate is achieved with the interrogation
time $T\simeq\pi/\Omega$ (see Fig.S2 in Supplementary Information), which is almost independent on the
dephasing rate as shown in Fig.\ref{fig:Pt-eta-de}(b). Our result shows that the dephasing will reduce
the measurement sensitivity to some extent but remains insignificant as long as the recombination
rate of the radical pair limits observation time rather than the NV decoherence rate. Hence, the achievable
sensitivity (e.g. under the influence of dephasing with a rate $\gamma=2\mbox{MHz}$), is still sufficient
to detect the effect of geomagnetic field orientation on a single-molecule radical pair reaction. We remark that the electric noise may also leads to an effective dephasing rate $\simleq 1\mbox{MHz}$ \cite{Kim15,Myers16} for a shallow NV center. Such an effect will reduce the sensitivity, see Fig.\ref{fig:Pt-eta-de}(b), but would not affect the feasibility of the present scheme. The present scheme works well for $k_{\mbox{eff}}$ within a certain range determined by $\Omega$ and $\gamma$, in particular it shall work for the typical value of $1/k_{\mbox{eff}}\sim 1\mu s$ \cite{Hor_16_AB} with a relatively high sensitivity ($\sim 1.53$ kHz Hz$^{-1/2}$).

{\it Conclusion.---} In summary, we propose a platform for the investigation of radical pair reactions
at the single-molecule level under ambient conditions using a quantum spin sensor in diamond. The present
scheme offers a highly sensitive way to measure the charge recombination rate of a radical pair reaction.
With a detailed analysis, we show that the achievable sensitivity is sufficient to detect the effect of
a magnetic field as weak as geomagnetic field on the radical pair reaction dynamics. The proposed
platform provides a potential route towards the study of single-molecule spin chemistry and quantum
coherent spin dynamics in radical pair reaction. The figure of merit of single-molecule measurement
allows the unambiguous identification of the relation between the properties of the radical pair
molecule and the function of a chemical compass, and therefore makes it possible to explore the role
of quantum coherence and entanglement in bio magnetic sensing. The hybrid design, when extended to
an ensemble of NV centers and radical pair molecules, can serve as an interesting construction to demonstration the principle of model chemical compass.

{\it Acknowledgements.---} We thank Prof. Ren-Bao Liu for helpful discussions and suggestions. J.-M.C
is supported by the National Natural Science Foundation of China (Grant No.11574103, 11690030, 11690032), the National
Young 1000 Talents Plan. H.-B. Liu is supported by China Postdoctoral Science Foundation Grant (Grant No. 2016M602274).
M.B.P. is supported by the ERC Synergy grant BioQ (Grant No. 319130).

\onecolumngrid
\renewcommand{\thefigure}{S\arabic{figure}}
\renewcommand{\theHfigure}{S\arabic{figure}}
\setcounter{figure}{0}
\renewcommand{\theequation}{S.\arabic{equation}}
\setcounter{equation}{0}

\newpage 

\section{Supplementary Information}

\begin{figure}[h]
\begin{centering}
\includegraphics[width=0.45\columnwidth]{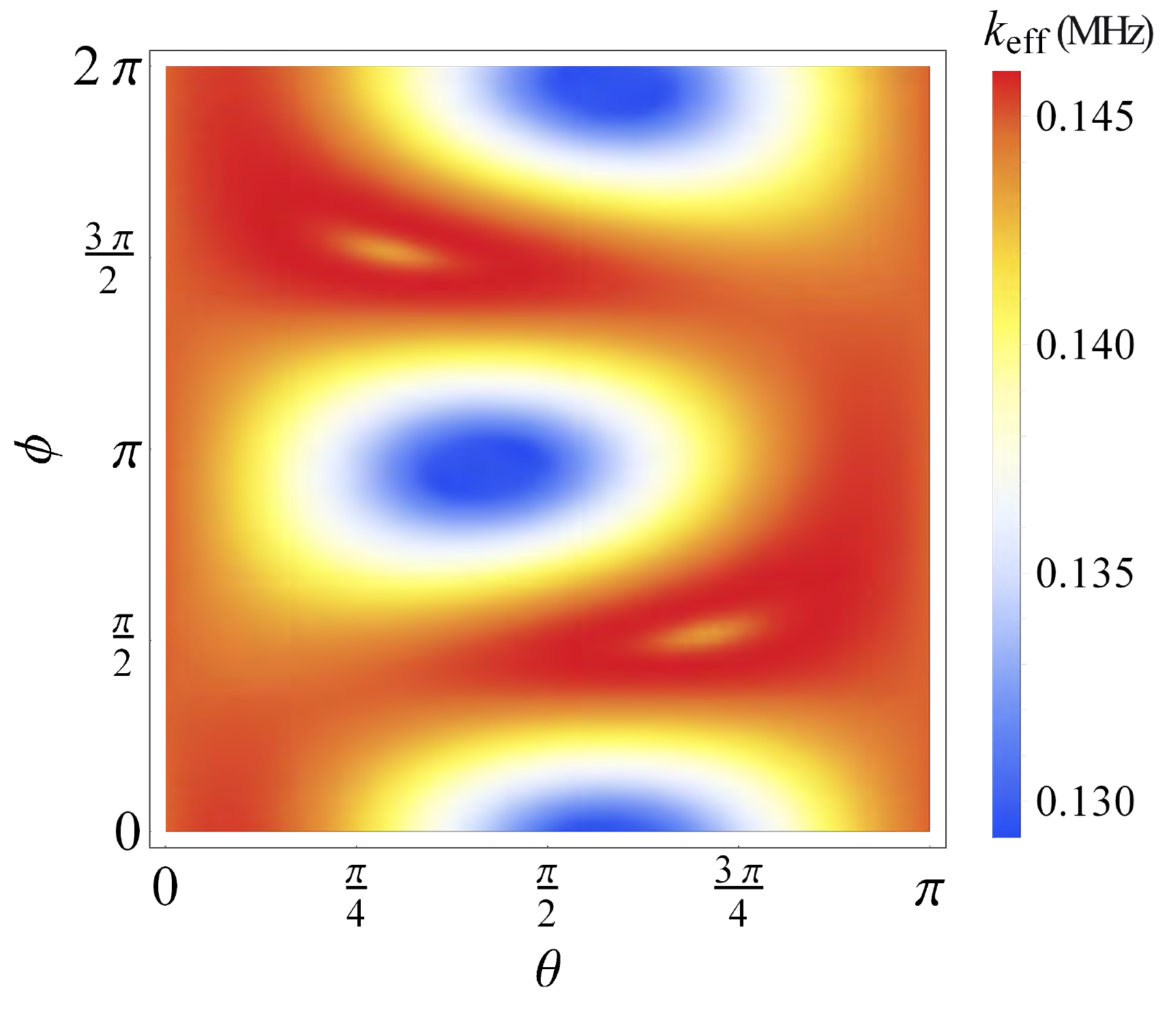}
\par\end{centering}
\caption{The effective charge recombination rate $k_{\text{eff}}$ as a function of the direction angles $(\theta,\phi)$ of the geomagnetic field. The parameters used here are $B_{0}=0.05$ mT, $k_{s}=0.02$ MHz and $k_{t}=0.2$ MHz. \label{fig:range} }
\end{figure}

{\it Magnetic effect on the effective charge recombination rate.---} We adopt the Krylov Space Methods \cite{Mol_03_SR} to numerically solve the master equations (4) and (12) in the main text. Without loss of generality, we only consider the nucleus H$^{6}$ with the dominant anisotropic hyperfine in the $\text{FH}{}^{\bullet-}$ radical \cite{Cin_03_CP_SI}. Besides all the parameters shown in the main text, the specific hyperfine parameters between the radical and the nucleus H$^{6}$ are shown in Table \ref{tab:H6} \cite{Cin_03_CP_SI}. To estimate the magnetic field effect on the radical pair reaction, we simulate the effective charge recombination rate $k_{\text{eff}}$ as a function of the direction of the geomagnetic field, denoted by $\theta$ and $\phi$, with different singlet and triplet reaction rates $k_{s}$ and $k_{t}$. As shown in Fig.\ref{fig:range}, the result shows about $10\%$ change in the effective charge recombination rate when a geomagnetic field changes its orientation with respect to the radical pair molecule, which agrees with the working principle of a radical pair mechanism based chemical compass.

\begin{table}[h]
\begin{centering}
\begin{tabular}{ccccc}
\hline
Nucleus & $A_{ii}$ & \multicolumn{3}{c}{Principal hyperfine axes}\tabularnewline
\hline
\hline
H$^{6}$ & -0.218 & -0.0362 & 0.2937 & 0.9552\tabularnewline
\hline
 & -0.202 & 0.7948 & 0.5879 & -0.1507\tabularnewline
\hline
 & -0.054 & -0.6059 & 0.7537 & -0.2546\tabularnewline
\hline
\end{tabular}
\par\end{centering}

\caption{The principal values of the hyperfine tensor $A_{ii}$ ($i=$ 1 to 3) and the principal hyperfine axes for nucleus H$^{6}$ \cite{Cin_03_CP_SI}. All hyperfine coupling parameters are in the unit of mT. \label{tab:H6}}
\end{table}

{\it Analytical solution of quantum master equation.---} The analytical form of the signal of the NV center spin sensor, which is shown as Eq.(9) in the main text, can be derived from the master equation {[}Eq.(4) in the main text{]}
\begin{eqnarray}
\dot{\rho} & = & -i\left[H,\rho\right]+\frac{k_{s}}{2}\left\{ 2L_{s}\rho L_{s}^{\dagger}-L_{s}^{\dagger}L_{s}\rho-\rho L_{s}^{\dagger}L_{s}\right\} \nonumber \\
 &  & +\frac{k_{t}}{2}\sum_{i=t_{0},t_{\pm}}\left\{ 2L_{i}\rho L_{i}^{\dagger}-L_{i}^{\dagger}L_{i}\rho-\rho L_{i}^{\dagger}L_{i}\right\}. \label{eq:ME-1}
\end{eqnarray}
We first express the above master equation in the charge separated and recombined state subspace, namely $\left|E\right\rangle =(1,0)^{\dagger}$, $\left|G\right\rangle =(0,1)^{\dagger}$, as follows

\begin{equation}
\left(\begin{array}{cc}
\dot{\rho}_{11} & \dot{\rho}_{12}\\
\dot{\rho}_{21} & \dot{\rho}_{22}
\end{array}\right)=-i\left(\begin{array}{cc}
[H_{E},\rho_{11}] & H_{E}\rho_{12}-\rho_{12}H_{G}\\
H_{G}\rho_{21}-\rho_{21}H_{E} & [H_{G},\rho_{22}]
\end{array}\right)+\left(\begin{array}{cc}
-k\rho_{11} & -\frac{k}{2}\rho_{12}\\
-\frac{k}{2}\rho_{21} & k\sum_{u=s,t_{0},t_{\pm}}l_{u}\rho_{11}l_{u}
\end{array}\right),
\end{equation}
where $l_{u}=\left|u\right\rangle \left\langle u\right|$, and $\sum_{u=s,t_{0},t_{\pm}}l_{u}=I$ has been used. After tracing out the degrees of freedom of the radical pair in $\rho_{11}$ and $\rho_{22}$, we obtain {[}Eq.(6) in the main text{]}
\begin{equation}
\left\{ \begin{array}{l}
\dot{\rho}_{E}=-i[H_{E},\rho_{E}]-k\rho_{E}\\
\dot{\rho}_{G}=-i[H_{G},\rho_{G}]+k\rho_{E}
\end{array}\right..
\end{equation}
Assuming $g_{e}\mu_{B}B_{0}\ll D$, $\theta=\frac{\pi}{2}$ and given the initial state $\rho_{E}(0)=\left|1\right\rangle \left\langle 1\right|$, we get the following result as
\begin{equation}
P^{E}(t)=\text{Tr}\left[\rho_{E}(t)\left|1\right\rangle \left\langle 1\right|\right] \simeq\frac{1+\cos(\Omega t)}{2}e^{-kt}.
\end{equation}
One can verify that $[H_{G},\rho_{G}]_{11}\simeq0$, so under the condition $k\ll k_{\perp}E_{\perp}$, we can obtain
\begin{equation}
P^{G}(t)\simeq k\int_{0}^{t}P^{E}(\tau)d\tau\simeq\frac{1-e^{-kt}}{2},
\end{equation}
which in turn allows us to get the result for the signal of the NV center spin sensor {[}Eq.(9) in the main text{]} as follows
\begin{equation}
P(t)=P^{E}(t)+P^{G}(t)\simeq\frac{1+\cos(\Omega t)e^{-kt}}{2}.\label{eq:Pt}
\end{equation}
The shot-noise limited measurement sensitivity is thus given by
\begin{equation}
\eta(T)\simeq \left|\frac{e^{k T} \sec (\Omega T ) \sqrt{1-e^{-2 k T} \cos ^2(\Omega T)}}{\sqrt{T}} \right|,
\end{equation}
which agrees well with the exact numerical result as shown in Fig.2(b) in the main text.\\

{\it Dephasing effect on the measurement sensitivity.---} An NV center spin in diamond suffers from dephasing noise due to the environment, e.g. nuclear spins or other impurities. The dynamics of the total system can be described by the following master equation {[}Eq.(11) in the main text{]}
\begin{eqnarray}
\dot{\rho} & = & \text{right-hand side of Eq.\eqref{eq:ME-1}}\nonumber \\
 &  & +\gamma\left\{ S_{z}\rho S_{z}-\frac{1}{2}S_{z}S_{z}\rho-\frac{1}{2}\rho S_{z}S_{z}\right\} .\label{eq:ME-dephasing-1}
\end{eqnarray}
Using similar techniques as above we can get
\begin{equation}
\left\{ \begin{array}{l}
\dot{\rho}_{E}=-i[H_{E},\rho_{E}]-k\rho_{E}+\gamma\mathcal{D}(\rho_{E},S_{z})\\
\dot{\rho}_{G}=-i[H_{G},\rho_{G}]+k\rho_{E}+\gamma\mathcal{D}(\rho_{G},S_{z})
\end{array}\right.,\label{eq:rho-eg-de-1}
\end{equation}
where $\mathcal{D}(\rho,S_{z})=S_{z}\rho S_{z}-\frac{1}{2}\left\{ S_{z}S_{z}\rho-\rho S_{z}S_{z}\right\} $. For $\theta=\pi/2$ and $g_{e}\mu_{B}B_{0}\ll D$, we find the solution of $\rho_{E}$ as follows
\begin{equation}
\rho_{E}=\left\{ \frac{1}{2}+\frac{e^{-\gamma t}}{2}\left[\cos(\Omega t)+\frac{\gamma}{\Omega}\sin(\Omega t)\right]\right\} e^{-kt},
\end{equation}
where $\Omega=\sqrt{(2k_{\perp}E_{\perp})^{2}-\gamma^{2}}$. Under the condition that $\gamma\ll k_{\perp}E_{\perp}$, we have
\begin{equation}
P ^{E}(t)\simeq\frac{1+e^{-\gamma t}\cos(\Omega t)}{2}e^{-kt}.\label{eq:p1e}
\end{equation}
In addition, one can also verify that $[H_{G},\rho_{G}]_{11}\simeq0$, $[\mathcal{D}(\rho_{G},S_{z})]_{11}=0$, therefore we get
\begin{equation}
P ^{G}(t)\simeq k\int_{0}^{t}P ^{E}(\tau)d\tau\simeq\frac{1-e^{-kt}}{2}.
\end{equation}
which leads to {[}Eq.(12) in the main text{]}
\begin{equation}
P(t)=P^{E}(t)+P_{1}^{G}(t)\simeq\frac{1+\cos(\Omega t)e^{-(k+\gamma)t}}{2}.\label{eq:p1t}
\end{equation}
\begin{figure}[t]
\begin{centering}
\includegraphics[width=0.5\columnwidth]{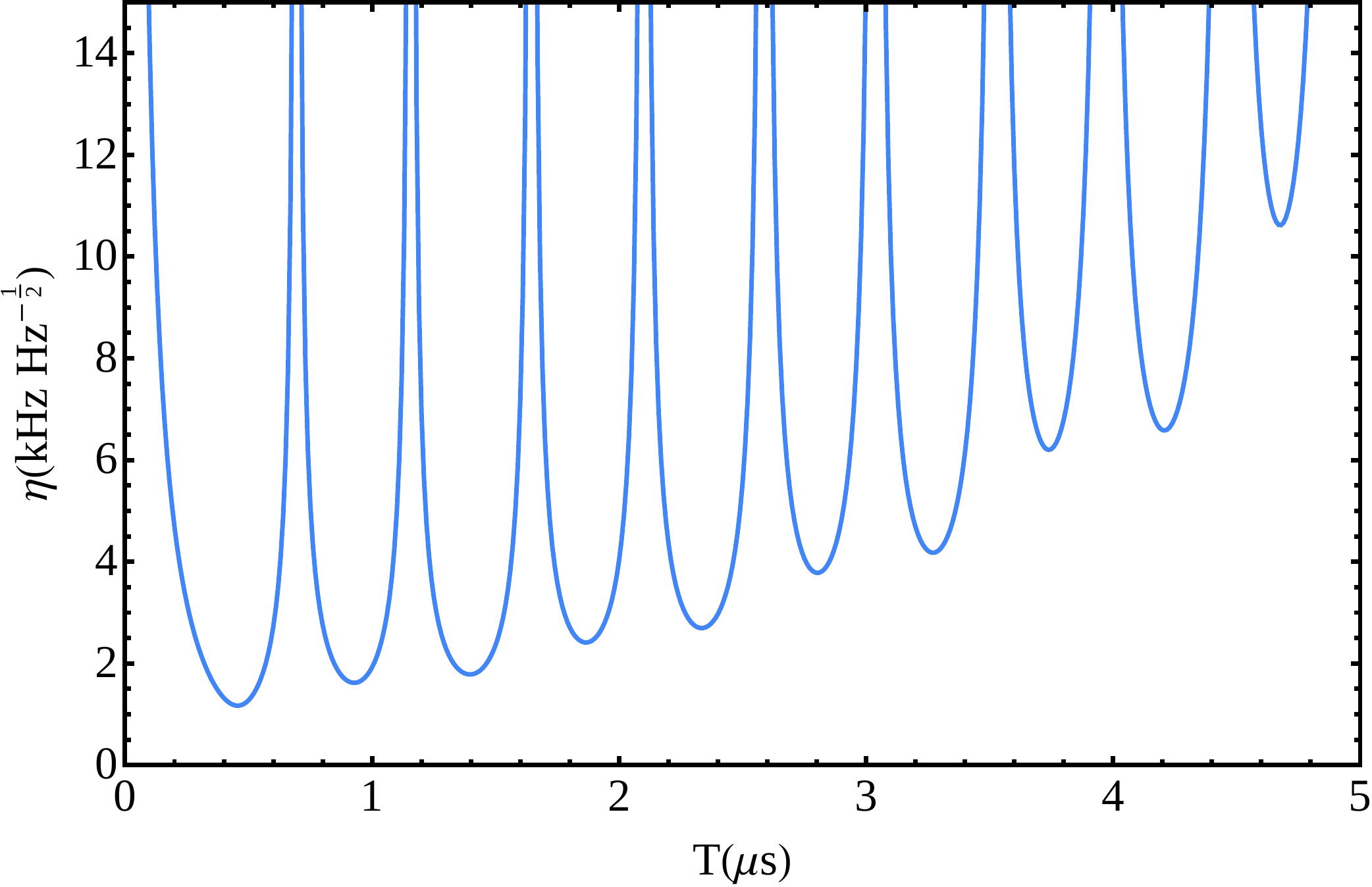}
\par\end{centering}
\caption{The sensitivity for the measurement of the charge recombination rate $k$ as a function of the interrogation time $T$. The best achievable sensitivity is estimated to be $\eta_{\text{op}}=1.17$ kHz Hz$^{-\frac{1}{2}}$ with $T=0.46\mu s$. The parameters are $B_{0}=0.05$ mT, $\theta=\frac{\pi}{2}$, $\phi=2.0$, $\gamma=0.5$ MHz and $k=k_{\text{eff}}=0.1425$ MHz. \label{fig:kappa} }
\end{figure}
The shot-noise limited measurement sensitivity is thus given by
\begin{equation}
\eta(T)\simeq \left|\frac{e^{(\gamma+k)T}\sec(\Omega T)\sqrt{1-e^{-2 (\gamma+k)T}\cos^{2}(\Omega T)}}{\sqrt{T}}\right|.
\end{equation}
The exact result of the measurement sensitivity from numerical simulation is shown in Fig.\ref{fig:kappa}. The result is similar to that in Fig.2(b) in the main text, although the sensitivity is slightly worse as Fig.2(b) takes
account of dephasing. In Fig.\ref{fig:Pt-eff-de}, we also verify that the dephasing will not influence the fitted value of the effective charge recombination rate.\\ \\
\begin{figure}[b]
\begin{centering}
\includegraphics[width=0.6\columnwidth]{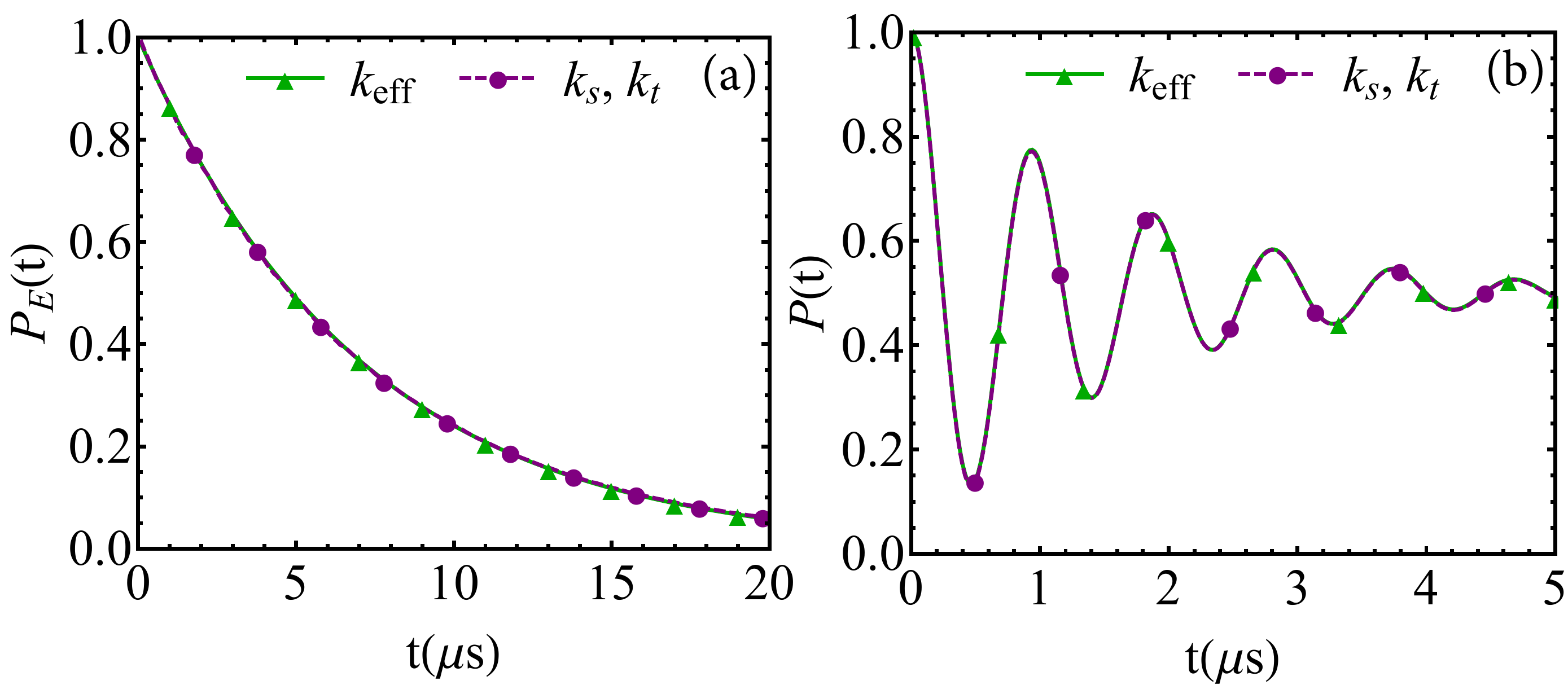}
\par\end{centering}
\caption{(Color online) The probability that the radical pair are in state $\left|E\right\rangle $ \textbf{(a)} and the NV center spin sensor is in state $\left|1\right\rangle$ \textbf{(b)} as a function of evolution time $t$. The green lines are the results using the effective charge recombination rate $k_{\text{eff}}$, while the purple dashed lines are that of $k_{s}$ and $k_{t}$. The parameters are $B_{0}=0.05$ mT, $\theta=\frac{\pi}{2}$, $\phi=2.0$, $k_{s}=0.02$ MHz, $k_{t}=0.2$ MHz, $\gamma=0.5$ MHz and $k=k_{\text{eff}}=0.1425$ MHz. \label{fig:Pt-eff-de} }
\end{figure}

For the more general case when $\gamma\ll k_{\perp}E_{\perp}$ is not satisfied, the solution in Eq.\eqref{eq:p1e} becomes as\begin{equation}
P ^{E}(t)\simeq\frac{1+e^{-\gamma t}C \cos(\Omega t-\alpha)}{2}e^{-kt}.\label{eq:p1e-1}
\end{equation}
where $C =\sqrt{1+\left({\gamma}/{\Omega}\right)^{2}}$ and $\alpha$ is defined by $\cos\alpha=\Omega/\sqrt{\Omega^{2}+\gamma^{2}}$. Therefore, the signal of the NV center spin sensor $P(t)$ can be obtained from
\begin{equation}
P (t)=P ^{E}(t)+k\int_{0}^{t}P ^{E}(\tau)d\tau
\end{equation}

\begin{figure}[h]
\begin{centering}
\includegraphics[width=0.9\columnwidth]{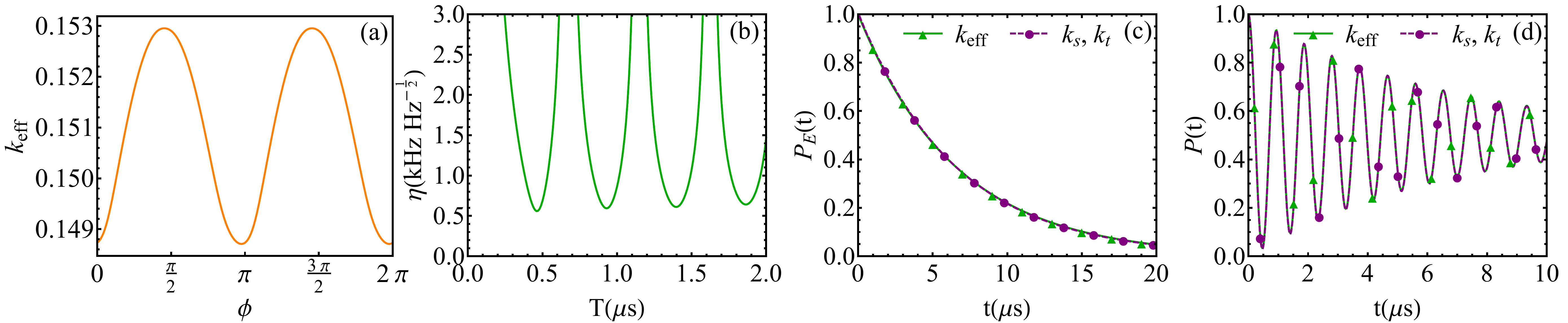}
\par\end{centering}
\caption{(Color online) {The influence of spin relaxation on quantum sensing of single-molecule radical pair reaction of Flavin radical ($\text{FH}{}^{\bullet-}$) and tryptophan radical ($\text{W}{}^{\bullet+}$)}. \textbf{(a)} The effective charge recombination rate $k_\text{eff}$ as a function of the direction angle $\phi$ when $\theta=\frac{\pi}{2}$ \textbf{(b)} The sensitivity for the measurement of the charge recombination rate $k$ as a function of the interrogation time $T$. The best sensitivity $\eta_{\text{op}}=0.56$ kHz Hz$^{-\frac{1}{2}}$ is achieved with $T=0.46\mu s$. \textbf{(c)} and \textbf{(d)} show the probabilities that the radical pair remains in the charge separated state $\left|E\right\rangle $ and the NV center is in the state $\left|1\right\rangle $ respectively as a function of the evolution time $t$. The Green lines are the results of $k_s=k_t=k_{\text{eff}}$, the purple dashed lines are that of $k_{s}$ and $k_{t}$ with different values. The spin relaxation rate is $\Gamma=0.1$ MHz, and the other parameters are the same as those of Fig.2 in the main text. \label{fig:Pt-eta-relax} }
\end{figure}

{\it The influence of spin relaxation of the radical pair.---} In the section, we estimate the effect of spin relaxation processes of the radical pair. Here, following Ref. \cite{Gau_11_PRL}, we describe the spin relaxation processes by a standard Lindblad formalism as 
\begin{eqnarray}
\dot{\rho} & = & \text{right-hand side of Eq.\eqref{eq:ME-1}}\nonumber \\
 &  & +\frac{\Gamma}{2}\sum_{i}\left\{ 2L_{i}\rho L_{i}^{\dagger}-L_{i}^{\dagger}L_{i}\rho-\rho L_{i}^{\dagger}L_{i}\right\} .\label{eq:ME-relax}
\end{eqnarray}
where the noise operators $L_i$ are $\sigma_x$, $\sigma_y$, $\sigma_z$ for each electron spin individually and for simplicity we use the same $\Gamma$ for all of them. The simulation results are shown in Fig.\ref{fig:Pt-eta-relax}. With the spin relaxation, the best sensitivity is slightly changed, however the angular sensitivity of the radical pair is indeed degraded. Therefore, if the relaxation is much faster, the detection of the radical pair reaction with the NV center spin sensor still works, but it becomes difficult for the NV center to detect the magnetic field's anisotropic effect on the radical pair, which itself is very small. 

\begin{figure}[h]
\begin{centering}
\includegraphics[width=0.4\columnwidth]{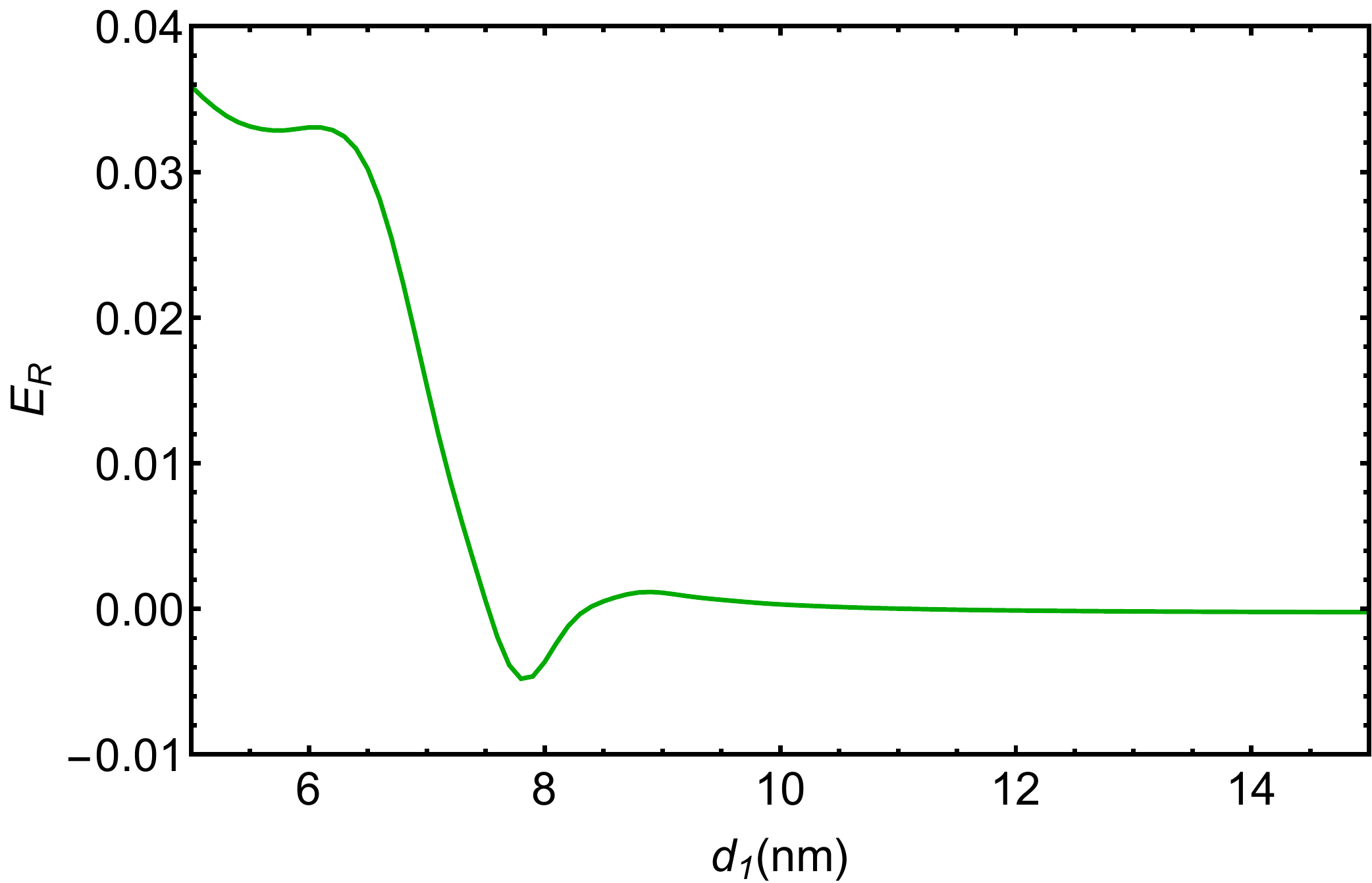}
\par\end{centering}
\caption{(Color online) Relative change ($E_R$) of the effective charge recombination rate $k_{\text{eff}}$ taking into account the dipolar spin-spin coupling between the radical pair and the NV center as a function of the depth of the NV center. The parameters used here are the same as those of Fig.2 in the main text except that the depth of the NV center $d_1$ is variable.\label{fig:dd} }
\end{figure}

{\it The influence of the dipolar spin-spin coupling between the radical pair and the NV center.---} In this section, the influence of the dipolar spin-spin coupling between the radical pair and the NV centre is carefully estimated for different values of the depth of NV center. As shown in Fig.\ref{fig:dd}, for the conditions that we assume in the main text, the influence is smaller than $5\%$ and drops rapidly with the increase of the NV center depth. 

\begin{figure}[h]
\begin{centering}
\includegraphics[width=0.35\columnwidth]{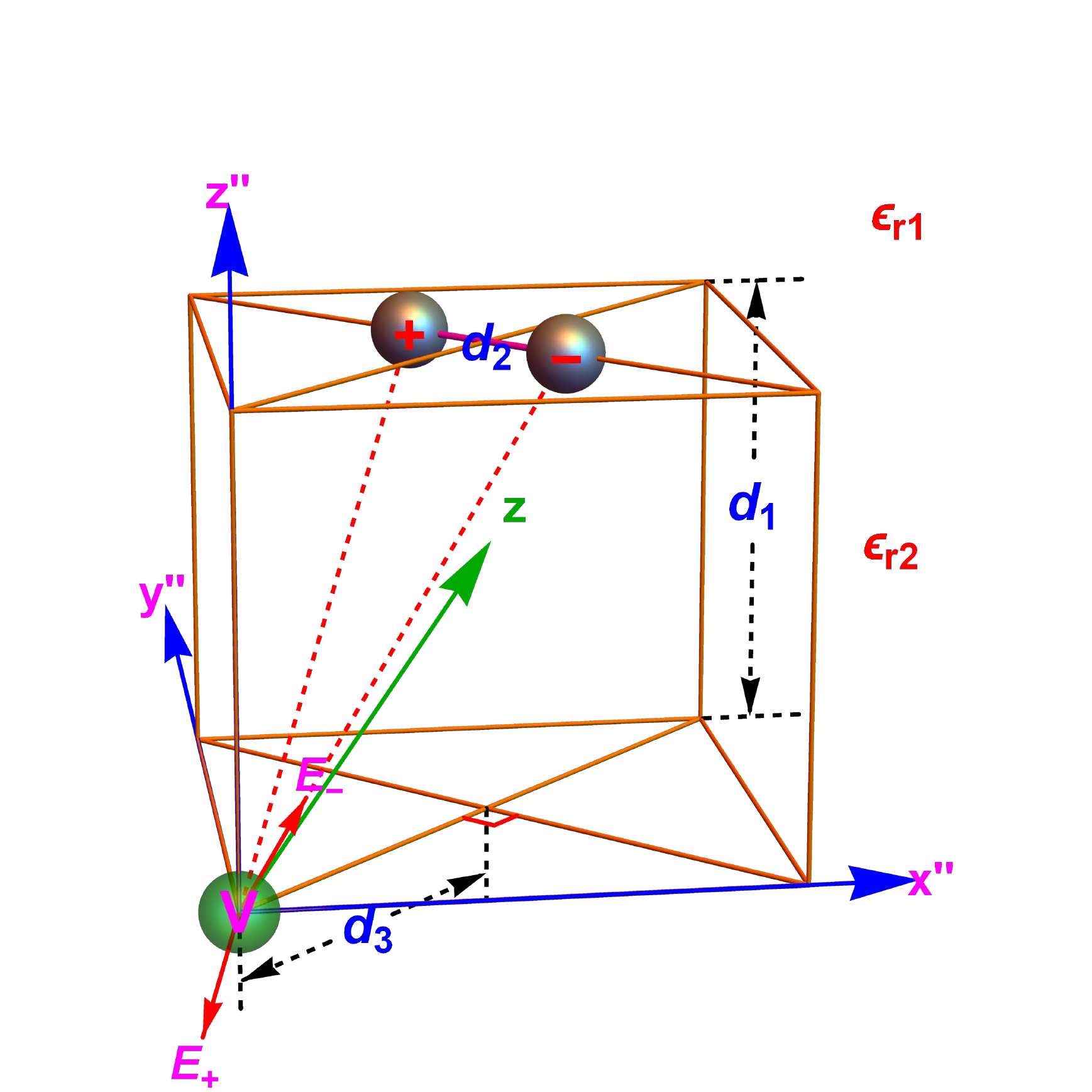}
\par\end{centering}
\caption{(Color online) The structure scheme used for the calculation of the electric field acting on he NV center. The green arrow indicates the  NV axis. The radical pair with a distance of $d_2$ between two radicals is located on the diamond crystal surface with a horizontal distance of $d_{3}$ from the NV center, and the depth of the NV center is $d_1$. \label{fig:E-dipole} }
\end{figure}

{\it Calculation of the electric field.---} The electric field generated by the radical pair is calculated according to the structure as shown in Fig.\ref{fig:E-dipole}. The radical pair are located on the medium interface of the air and the diamond crystal, whose relative permittivities are $\epsilon_{r1}$ and $\epsilon_{r2}$ respectively. As an example, we assume that the radical pair is $d_{2}=2$ nm apart with a horizontal distance about $d_{3}=4$ nm to the NV center, and the depth of the NV center is $d_1=5$ nm. The electric field is calculated by the method of electrical images as
\begin{equation}
\vec{E}=\frac{q\hat{\vec{r}}_{+}}{4\pi\epsilon_{0}r_{+}^{2}}\left(\frac{2}{\epsilon_{r1}+\epsilon_{r2}}\right)-\frac{q\hat{\vec{r}}_{-}}{4\pi\epsilon_{0}r_{-}^{2}}\left(\frac{2}{\epsilon_{r1}+\epsilon_{r2}}\right)
\end{equation}
\begin{figure}[h]
\begin{centering}
\includegraphics[width=0.35\columnwidth]{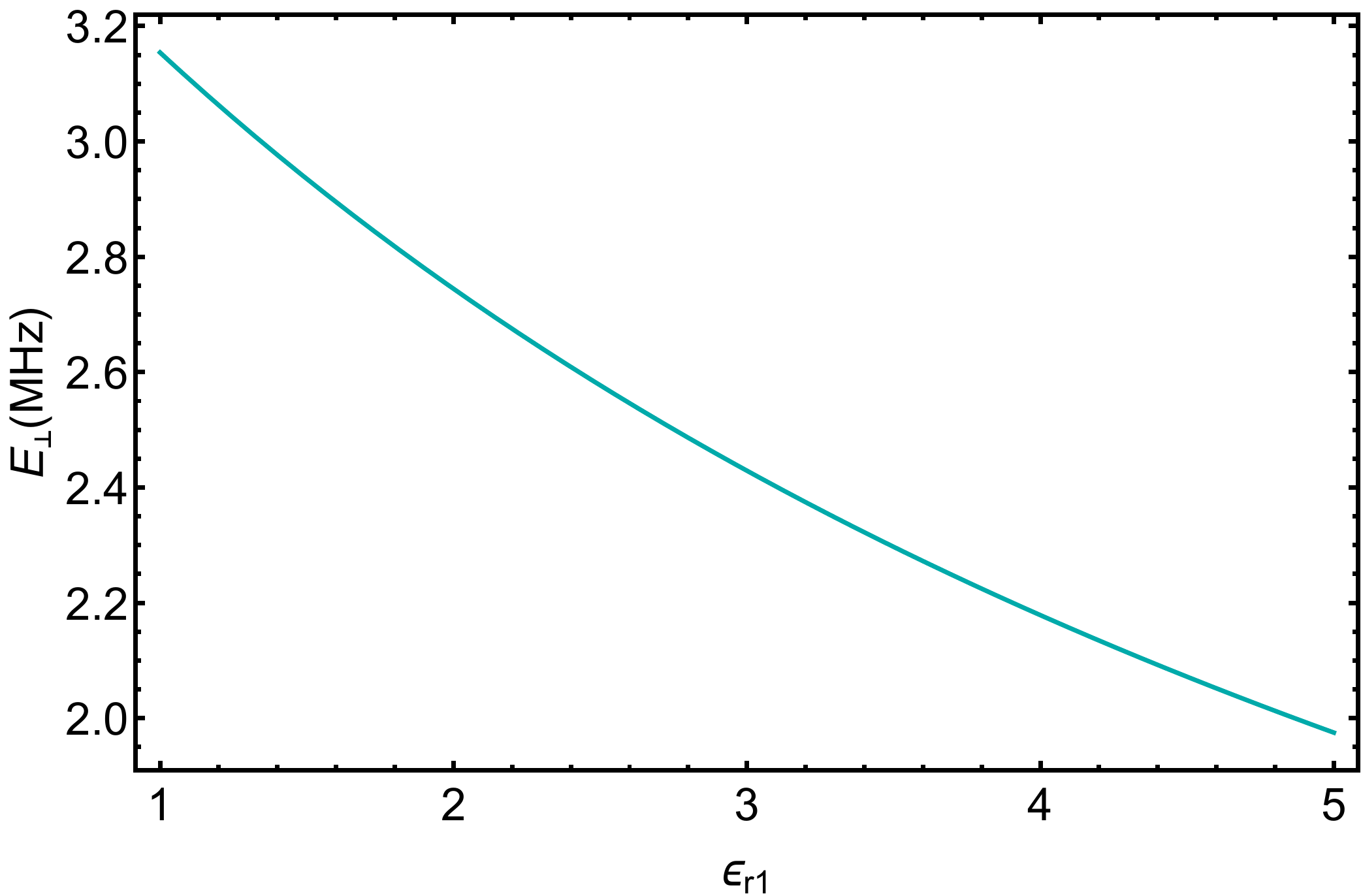}
\par\end{centering}
\caption{(Color online) The transverse electric field as a function of the permittivity of the medium $\epsilon_{r1}$. \label{fig:er1} }
\end{figure}
where $q$ is charge magnitude of the electron, $\epsilon_0$ is the vacuum permittivity, $r_+$ ($r_-$) is the distance between the NV center and the positive (negative) charge, $\hat{\vec{r}}_{+}$ ($\hat{\vec{r}}_{-}$) is the unit vector from the positive (negative) charge to the NV center. In our scheme, the main effect of the electric field comes from its component ($E_{\perp}$) along the direction that is in perpendicular to the NV axis which connects the nitrogen atom and the vacancy site, accounting for the relative permittivity of diamond ($\epsilon_{r2}=5.7$) and air ($\epsilon_{r1}=1.0$), the calculated transverse electric field is $E_{\perp}=3.15$ MV/m, which is used in our paper. \\

{\it Decoupling pulses.---} In the absence of decoupling pulses, the evolution of the NV center is $U_{0}=e^{-iH\tau}$, where $H$ is either $H_{NV}$ or $H_{NV}\vert_{ E=0 }$. The effect of the transverse component of the magnetic field (included in $H$ as $\gamma B_x S_x$, $\gamma B_y S_y$) can be eliminated to the second order of $\tau$, where $\tau$ denotes the time interval between pulses, by the controlled evolution as follows
\begin{eqnarray*}
U & = & U_{0}U_{z}U_{0}U_{0}U_{z}U_{0}\\
 & = & U_{0}\left(U_{z}U_{0}U_{z}\right)\left(U_{z}U_{0}U_{z}\right)U_{0}\\
 & = & e^{-i4\tau\left[H_{\O}+O(\tau^{2})\right]}
\end{eqnarray*}
where $U_{z}=\left|1\right\rangle \left\langle 1\right|-\left|0\right\rangle \left\langle 0\right|+\left|-1\right\rangle \left\langle -1\right|$, $H_{\O}=(D+k_{\parallel}E_{z})(S_{z}^{2}-2/3)-k_{\perp}E_{x}(S_{x}^{2}-S_{y}^{2})+k_{\perp}E_{y}(S_{x}S_{y}+S_{y}S_{x})$. \\

\begin{figure}[h]
\begin{centering}
\includegraphics[width=1.0\columnwidth]{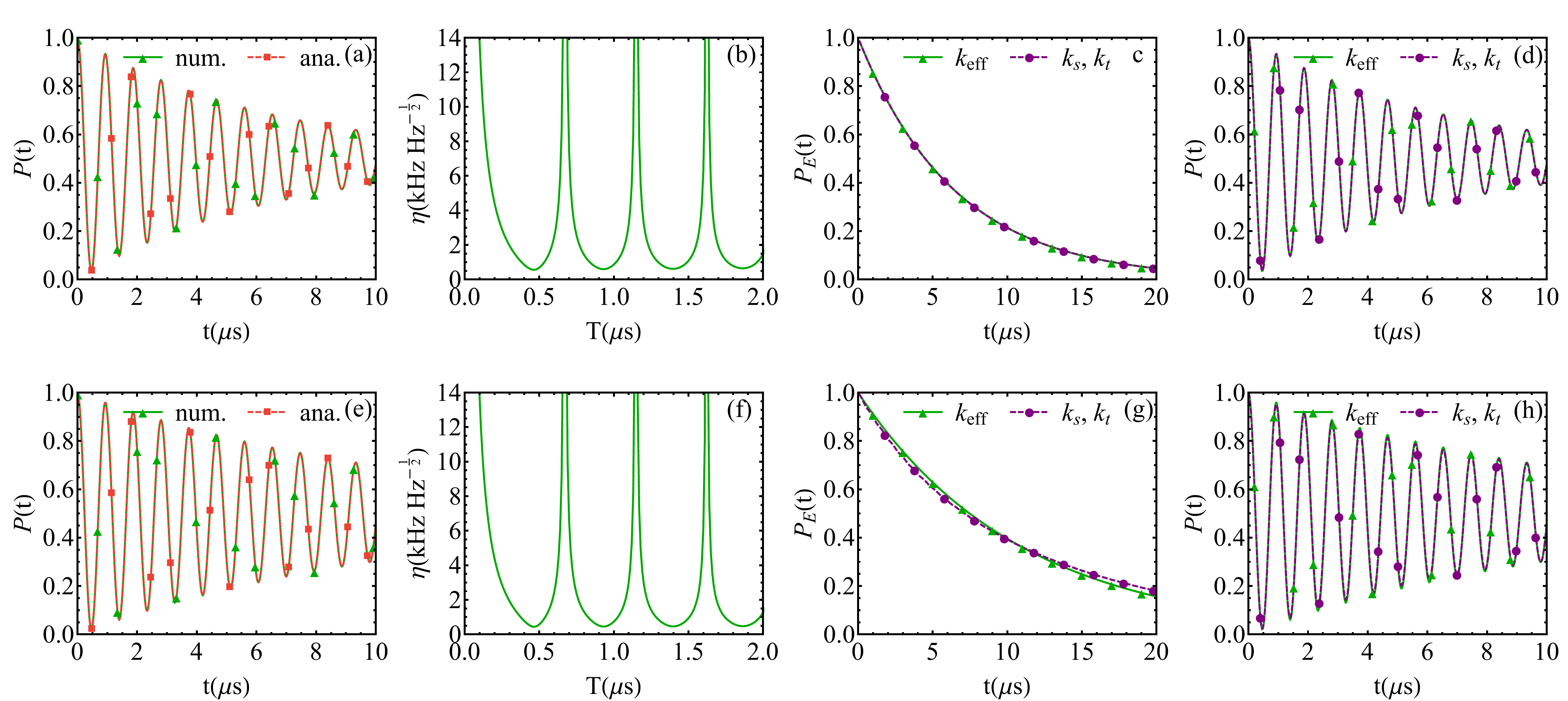}
\par\end{centering}
\caption{(Color online) {Quantum sensing of single-molecule radical pair reaction of Flavin radical ($\text{FH}{}^{\bullet-}$) and tryptophan radical ($\text{W}{}^{\bullet+}$)}. \textbf{(a-d)} are the results when including H$^5$ in $\text{FH}{}^{\bullet-}$, \textbf{(e-h)} are the ones when including N$^5$ in $\text{FH}{}^{\bullet-}$. \textbf{(a)} and \textbf{ (e)}: The effective charge recombination rate $k_\text{eff}$ as a function of the direction angle $\phi$ of the magnetic field when $\theta=\frac{\pi}{2}$. \textbf{(b)} and \textbf{(f)}: The sensitivity for the measurement of the charge recombination rate $k$ as a function of the interrogation time $T$. The best sensitivity $\eta_{\text{op}}=0.563$ kHz Hz$^{-\frac{1}{2}}$ is achieved with $T=0.46\mu s$ in \textbf{(b)}, the results of \textbf{(f)} are $\eta_{\text{op}}=0.436$ kHz Hz$^{-\frac{1}{2}}$ and $T=0.46\mu s$. \textbf{(c)}\textbf{(g)} and \textbf{(d)}\textbf{(h)} show the probabilities that the radical pair remains in the charge separated state $\left|E\right\rangle $ and the NV center is in the state $\left|1\right\rangle $ respectively as a function of the evolution time $t$. The Green lines are the results of $k_s=k_t=k_{\text{eff}}$, the purple dashed lines are that of $k_{s}$ and $k_{t}$ with different values. The other parameters are the same with Fig.2 in the main text.\label{fig:h5n5} }
\end{figure}

{\it Radical pair with different nuclei.---} In the main text, we choose H$^6$ as the unique hyperfine interaction for the simplicity of calculation. Of course, the realistic model shall include multi nuclear spins, which inevitably requires very large computing resource. Usually the number and specie of the nuclear spins in the radical pair will affect the change of the effective charge recombination rate in response to the external magnetic field, but the best achievable sensitivity for the measurement of the effective recombination rate using the NV center spin sensor is slightly affected. We show the results with  the choice of different nucleus in Fig.\ref{fig:h5n5}(a-d) (H$^5$) and Fig.\ref{fig:h5n5}(e-h) (N$^5$), which agree well with our arguments.

\begin{figure}[h]
\begin{centering}
\includegraphics[width=0.6\columnwidth]{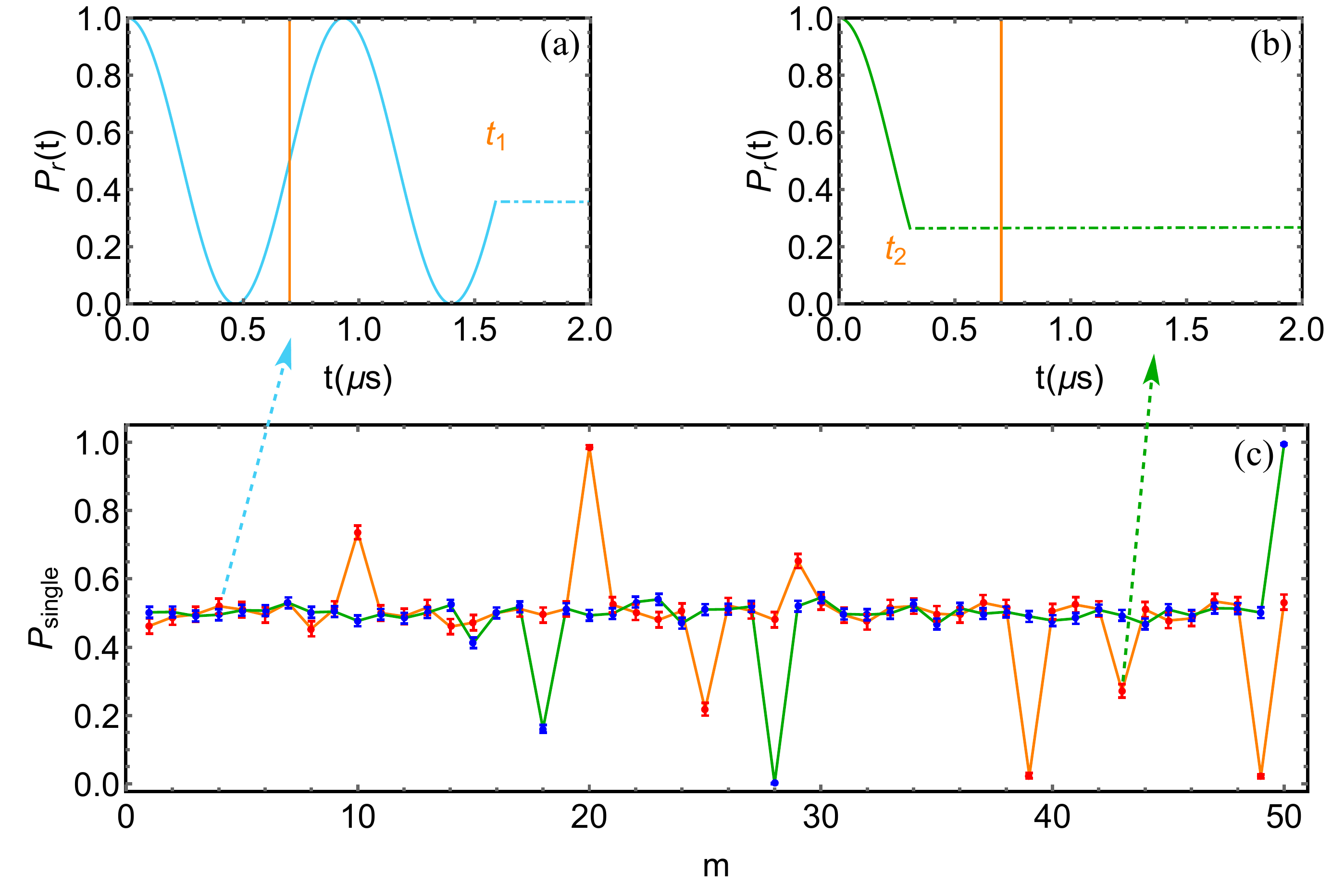}
\par\end{centering}
\caption{(Color online) {Quantum sensing of single-molecule radical pair reaction of Flavin radical ($\text{FH}{}^{\bullet-}$) and tryptophan radical ($\text{W}{}^{\bullet+}$)} using the single-shot readout. \textbf{(a)} or \textbf{(b)} demonstrates two measurement trajectory of a single-molecule detection process where the electric field disappears at $t_{1}$ ($t_{1}>t_{m}$) or $t_{2}$ ($t_{2}<t_{m}$), the two orange lines indicates that the measurements occur at $t_{m}=0.7\mu s$, the probabilities of these two kinds of events are $\int_{0}^{t_{m}}ke^{-kt}dt=0.095$ and $\int_{0}^{t_{m}}ke^{-kt}dt=0.905$ respectively. \textbf{(c)} shows the signal $P_{single}$ for Rabi oscillation measurements from the $m$th single-molecule detection event. Each data point represents is a simulation by $N=500$ repetitions in a one-time single-shot readout. The error bars are calculated as $\sqrt{\left(P_{r}(\tau)-P_{r}(\tau)^{2}\right)/N}$, where $\tau=t_{m}\:or\:t$, depending on if $t>t_{m}$ or not. The parameters are the same with Fig.2 in the main text.\label{fig:ss} }
\end{figure}

{\it The measurement trajectory of single-molecule detection.---}For single-molecule detection, the radical is either present or absent, so the electric field experienced by the NV center would drop to zero after the radical pair has recombined. This kind of processes are showed in Fig.\ref{fig:ss} (a) and (b), when the electric field disappears, the Rabi oscillation ($P_{r}(t)=(1+\cos(\Omega t))/2$) stops. In addition, when an individual radical pair is formed the nuclear spins coupled with the radical pair has a specific nuclear spin configuration, the average of which over a large number of detections corresponds to a thermal equilibrium nuclear spin state. In the present scheme, it requires to determine the state probability of the NV center spin sensor, it is possible to achieve this using a technique called single-shot readout \cite{Jia_09_S, Neu_10_S}, which facilitates the measurement of single-molecule detection. Here we show how the average of such single-molecule detection data leads to the ensemble average result that is the same as Fig.2 in the main text. In a single-molecule detection, the probability density that the electric field disappear at time $t$ is $ke^{-kt}$, where $k$ is the effective charge recombination rate. If the measurement occurs before $t$, we will get $P_{r}(x)$ (e.g. Fig.\ref{fig:ss} (a), where $x$ is the measurement time), otherwise the result is $P_{r}(t)$ (e.g. Fig.\ref{fig:ss} (b)). So taking into account a random nuclear spin configuration, the probability that NV center is in state $\left|1\right\rangle$ at time x can be calculated as
\begin{eqnarray}
P(x) & = & \frac{1}{2}\left(\int_{0}^{x}P_{r}(t)k_{1}e^{-k_{1}t}dt+P_{r}(x)\int_{x}^{\infty}k_{1}e^{-k_{1}t}dt\right)+\frac{1}{2}\left(\int_{0}^{x}P_{r}(t)k_{2}e^{-k_{2}t}dt+P_{r}(x)\int_{x}^{\infty}k_{2}e^{-k_{2}t}dt\right)\\
 & \simeq & k\int_{0}^{x}P_{r}(t)e^{-kt}dt+P_{r}(x)e^{-kx}
\end{eqnarray}
where $k_{1}$ and $k_{2}$ are the effective charge recombination rates corresponding to different nuclear spin configurations. This equation can recover the result of equation\eqref{eq:Pt}, so as showed in Fig.\ref{fig:ss} (c), if we repeat the detection for many times, the average will converge to $P(x)$.

\begin{figure}[h]
\begin{centering}
\includegraphics[width=0.4\columnwidth]{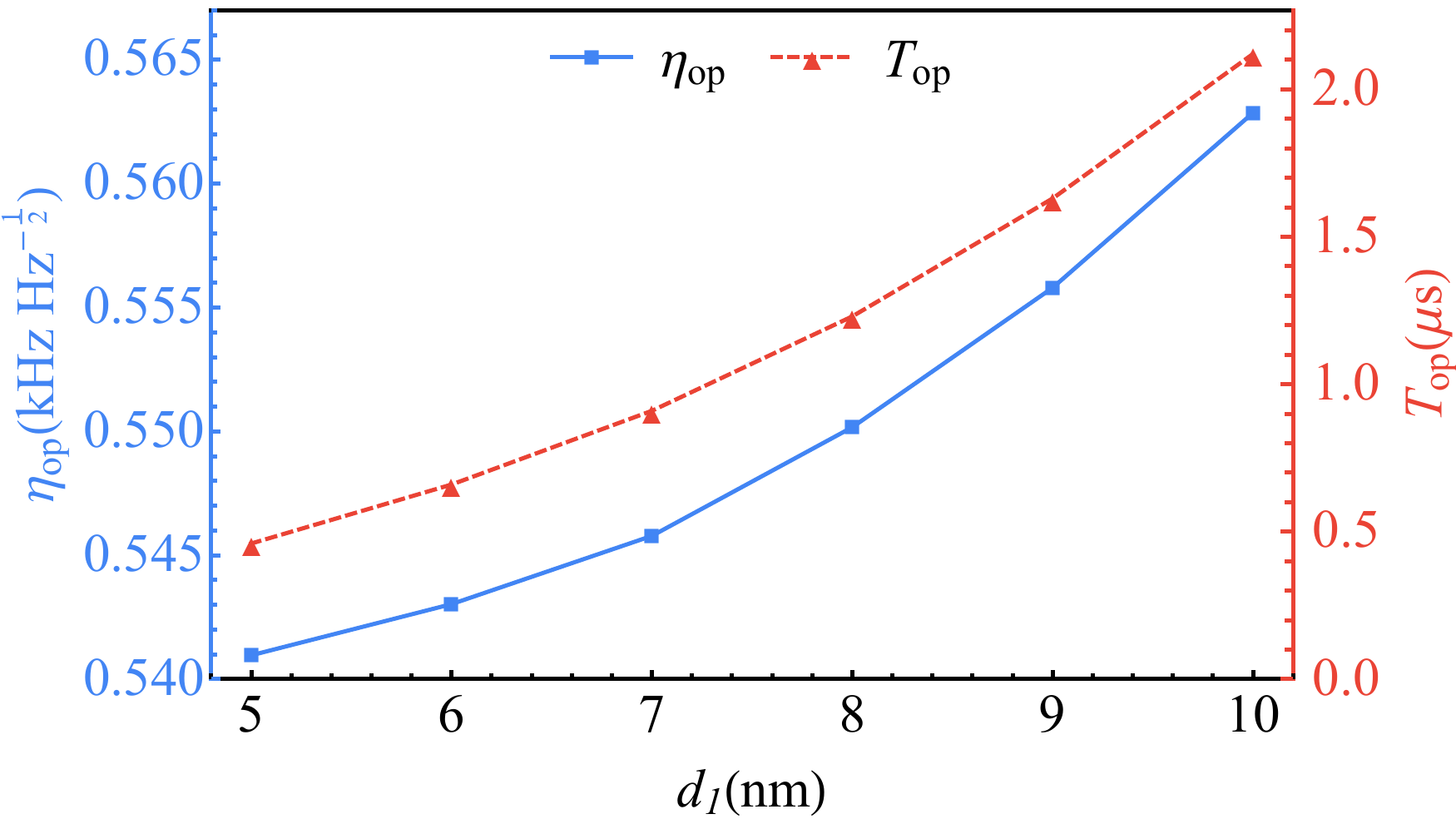}
\par\end{centering}
\caption{(Color online) The estimated achievable sensitivities as a function of the NV-molecule distance. The other parameters are the same with Fig.2 in the main text.\label{fig:depth} }
\end{figure}

{\it Achievable sensitivities for larger NV-molecule distances.---} In practice, radical pair which is sitting inside a molecule may have a larger distance from the NV center. To estimate the effect of the NV-molecule distance on the measurement sensitivity, in Fig.\ref{fig:depth} we show the estimated achievable sensitivities and the corresponding optimal measurement times for larger NV-molecule distances. The sensitivity becomes worse as the distance increases, the change of which is however not very significant for the distance below 10nm, although the measurement time becomes larger as the distance increases. This would thus require a longer coherence time of the NV center spin.

\end{document}